# On the Coupling of Pressurized Flow and Elastic Expansion of Artificial Rocks

Arnold Bachrach & Yaniv Edery

Faculty of Civil and Environmental Engineering, Technion, Haifa, Israel.

Abstract

Pressurized fluid injection into underground rocks occurs in applications like carbon sequestration, hydraulic fracturing, and wastewater disposal, and may lead to human-induced earthquakes and surface uplift. The fluid injection raises the pore pressure within the porous rocks, while deforming them, yet this coupling is not well understood as experimental studies of rocks are usually limited to postmortem inspection and cannot capture the complete deformation process in time and space. We investigate injection-induced deformation of a unique rock-like transparent medium mimicking the deformation of sandstone, yet under low pressure. By incorporating within this artificial rock fluorescent microspheres we capture its internal deformation in real time during the pressurized flow. We then modify the theory of poroelasticity to model accurately and without any fitting parameters the internal elastic deformations, hence providing a physical mechanism for the process. Moreover, our results demonstrate and validate the underling assumptions of the poroelastic theory for fluid injection in rock-like materials. Our results are relevant for understanding human-induced earthquakes and injection induced surface uplift, as they decouple the role of the pressurized flow from the rock deformation through the poroelastic theory.

Introduction

In recent years the coupling of flow, pressure and deformation in porous media proves to be seminal to our understanding of fluid injections into the underground, as is the case for carbon sequestration[1,2], hydraulic fracturing[3], enhanced geothermal energy production[4], and wastewater disposal[5–7]. The fluid injection raises the pore pressure within the underground rocks, leading to human-induced earthquakes[5–10] and surface uplift[5,7,11]. While human-induced earthquakes are attributed to fault reactivation[8–10], surface uplift is a phenomenon that can be attributed to elastic expansion of the underground rocks[5,12,13]. Moreover, the elastic response of the medium to fluid injection can induce underground stresses and facilitate fault reactivation[14,15]. Early observations on the coupling of flow, pore-pressure, and elasticity have been documented by King (1892)[16], as he measured water-level fluctuations in wells due to passing trains [14,15]. It was later shown by Jacob (1939)[17] that the weight of a train can elastically compress the underground aquifer, hence raising its pore water pressure and elevating the surrounding wells water-level. These observations were accompanied by comprehensive scientific study, mainly in the context of soil consolidation[18–20] and elastic storage in a confined aquifer[21–23], formulating the theory of Poroelasticity.

Originally, the poroelastic theory was developed for the case of fluid outflow from the porous medium- weather by fluid extraction[21,23] or by loading[18,19], hence, experiments that prove and demonstrate the validity of this theory for fluid injection scenarios are rare. Moreover, injection experiments on rocks are usually limited to postmortem inspection [24,25] and cannot capture the complete spatial and temporal dynamics of the deformation process. Although experiments on soft or loosely consolidated materials provide meaningful insights on injection induced deformation[26–28], the materials used in these experiments are very different from real rocks by their solid internal structure, mechanical behavior, and permeability.

In this study, we will investigate injection-induced poroelastic expansion of a unique rock-like transparent medium that mimics the deformation of sandstone under low pressure. Our medium undergoes all the deformation range of real rock, from elastic to plastic, while allowing us to capture the whole internal displacement field. We will show that the mechanism behind steady-state fluid injection leads to non-uniform expansion of the medium, a phenomenon that we fully model using the theory of poroelasticity[18–20,29].

Methods

To investigate the coupling of pressurized flow and deformation, we developed a transparent artificial rock by sintering PMMA (Polymethyl Methacrylate) beads with a mean diameter of 78 microns (Appendix A, figure 4a, b). We chemically sinter the beads inside a self-made PMMA chamber using an Acetone mixture that dissolves the edges of the beads and subsequently drains the mixture, allowing the beads to solidify together into a rock-like porous material (figure 1a, b; Appendix A figure 4c, d). At the injection point, the sample is fixed by an epoxy glue to the flow-cell wall while the outlet edge remains free (figure 1a). The artificial rock porosity ($n = 0.43$) is measured from micro-CT scan of the sample (Appendix A figure 4d). The artificial rock permeability, calculated by the Kozeny-Carman equation[30,31] ($k = 5.85 X 10^{-12}\ m^2$), corresponds to a high-permeability sandstone[32]. To track the local deformation, we incorporate and solidify 1.2% fluorescent Polyethylene microspheres (106-125 microns in diameter) within the sample (figure 1b). The tracking is achieved by saturating the sample with oil (Cargille immersion liquid) that has a matching refractive index as the PMMA (RI=1.49), transforming the sample from opaque to transparent, apart from the fluorescing microspheres which are excited by 460 nm LED light and emit at a range of 580-700 nm, and filtered by a long-pass 625 nm filter (figure 1c).

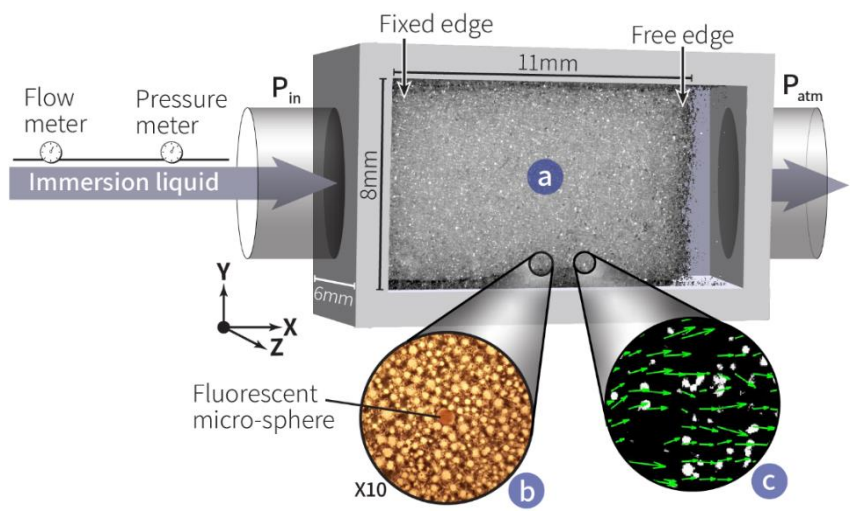

Figure 1. Experimental set-up. a: The artificial rock-like sample within the flow-cell, where the injection point is fixed to the flow-cell wall while the outlet edge remains free. b: The sintered PMMA beads with the fluorescent micro-spheres forming the artificial rock, as seen under optical microscope. c: Tracking and analysis of the filtered fluorescent microspheres by an ultra-high-speed camera and PIV software[33], respectively, which provide the whole displacement field within the sample.

The oil flow through the artificial rock is driven by a pressure difference using a pressure pump (Fluignt-LU-FEZ-7000) as we monitor the inlet pressure and fluid discharge by a pressure-sensor (Fluignt-EIPS7000) and flowmeter (Fluignt-FLU-XL), respectively. This pressure difference increases by 70 mbar/sec as we track the deformation by the fluorescent microspheres movement using a high-resolution (4Mpx) ultra-high-speed camera (Phantom-v2640) at a rate of 100 frames per second at 12 bits. A Particle Image Velocimetry (PIV) software (PIVlab 2.50)[33] provides the whole displacement field within the sample for each timestep by analyzing the fluorescent microsphere's joint movement (figure 1c; Appendix B, figure 5). We calculated the PIV accuracy as 0.57 microns (0.04 pixels) by applying the analysis on the system without any pressure difference, where the displacement of the fluorescent beads is zero, to measure the imaging noise.

## Results

The embedded fluorescent micro-sphere movement in the artificial rock allows us to calculate the displacements within the sample using the PIV software[33] for each inlet pressure that drives the flow. By addressing the displacement of the sample's free edge, a pseudo-stress-strain curve as in a rheological test can be derived (figure 2a). In figure 2a, the $x$ axis is the sample's free edge displacement averaged over the $y$ axis of the sample, and normalized by the sample's initial length. Thus, providing the overall strain of the sample. The $y$ axis in figure 2a is the inlet (gauge) pressure, representing the stress in the pseudo-stress-strain curve. Looking at the curve, we identify a trend similar to that of a pulled rock in a tensile test: linear extension for pressures of up to 0.084 MPa, followed by a non-linear extension. Moreover, the normalized linear extension of our sample is the same as a pulled sandstone strain in a pulling test, while having the same scale for the non-linear extension (figure 2b)[34]. We verify the pseudo-stress-strain curve transition from elastic to plastic by a cyclic pressure test (figure 2c), where we apply cycles of pressure increase followed by a pressure decrease, with an increasingly higher pressure for each cycle. As seen in figure 2c, the strain at cycle 1, reaching about 0.1 MPa, is completely reversible, meaning that the deformation is elastic and non-dissipating. However, in the following cycles, we can see a larger and larger remnant strain, manifested in the hysteresis of the cycle, meaning that indeed a plastic component is added to the deformation.

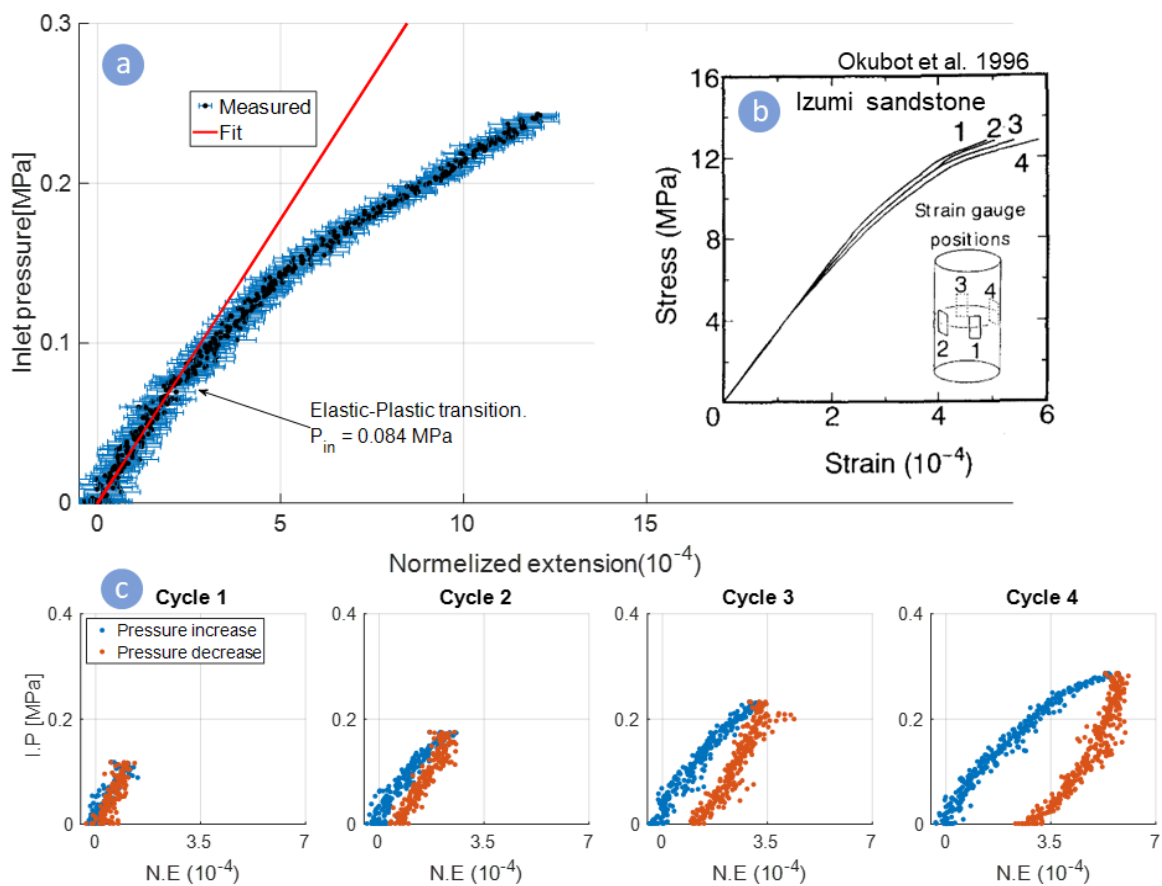

*Figure 2. Pseudo stress-strain curves. a: The strain is the total extension normalized by the sample's initial length, and the stress is the measured inlet pressure. The curve follows the same transition between elastic to plastic extension observed in a pulled rock. The error of the pressure sensor (1.6e-3 to 4e-3 MPa) is negligeable compared to the PIV error (0.6 microns). b: Comparison of our results to a sandstone pulling test done by Okubot et.al (1996) [34]. The extension of our sample is similar not only in the general trend but also in the scale of strain. c: Cyclic pressure tests, where the end pressure applied on the sample is increasing on every cycle, and subsequently decrease. The strain at cycle 1 reaches about 0.1 MPa is completely reversible, indicating an elastic non-dissipating response, however, the following cycles with higher end pressure are not reversible, showing a remnant plastic strain manifested by each cycle hysteresis. I.P: inlet pressure, N.E: normalized extension.*

Our artificial rock simulates real rock deformation by moving from the elastic to the elasto-plastic regime. However, unlike opaque rocks, it allows us to quantify the coupling between the pressurized flow and the deformation by analyzing the internal local displacement as the pressure drops from inlet to outlet. We analyze the internal displacement of the experiment in figure 2a by calculating the mean displacement over the sample's $y$ axis for each 224 microns along $x$ (see more details in Appendix B). Figure 3 shows the measured displacement along the sample for three different inlet pressures from the same experiment. As can be seen, the increase in displacement along $x$ is not distinctly linear as one would expect from the analogy to a pulling test, suggesting a non-uniform strain distribution. To understand the measured results, we merged Hooke's law, Darcy's law, and Terzaghi's concept of effective stress[19,20,36], which coincide with Biot's theory of poroelasticity[18] for incompressible fluid and grains.

The pore Reynolds number along the experiment has been calculated to be <0.1, hence the flow through the porous medium should be governed by Darcy's law[31,37], where gravitational forces and momentum are negligible compared to the pressure difference. Our boundary conditions allow flow just in the $x$ direction, hence we can use the one-dimensional Darcy's law that relates the pore pressure drop ($\frac{dp}{dx}$) with the fluid flux ($q$), through the permeability ($k$) and the dynamic viscosity of the fluid ($\mu = 11.7 \ mPa*s$):

$$\frac{dp}{dx} = -\frac{\mu}{k} q .  \qquad (1)$$

Due to the small elastic deformations of the artificial rock (figure 2a), the change in porosity measured from the sample's elongation is negligible (<0.1%), hence we will treat the permeability as a constant, which means that while the pore-pressure strains the medium, the deformation has a negligible effect on the flow. We further assume that the response time for each inlet pressure is instantaneous due to the small sample size, so for each inlet pressure the system is at steady state and the flux is constant along the sample ($\frac{dq}{dx} = 0$).

Since $q, k$ and $\mu$ are assumed to be constants the pore-pressure will drop linearly along the sample. Under the conditions $p(x = 0) = p_{injection}$ and $p(x = L) = p_{atm}$ for a sample at length $L$, the pore pressure along the sample will be:

$$p(x) = -\frac{p_{in}}{L}x + p_{in}. \qquad (2)$$

Although the sample is extending like in a pulling test (figure 2), the boundary conditions are that of Oedometric test[38,39], where the sample is compressed vertically while confined laterally. This is due to the tensional stress caused by pore pressure in all directions[18–20,36]. That is, our set-up allows the sample to expand only in the $x$ direction, while both $y$ and $z$ directions are constrained by the flow-cell walls for expansion (see figure 1 for orientation), so there is one direction of stress ($\sigma_x$) and free strain ($e_x$) and two other directions of no strain ($e_y = e_z = 0$) and reaction stresses ($\sigma_y, \sigma_z$), just like in Oedometric test. Using these boundary conditions with Hook's law for isotropic elastic body, while addressing the presence of fluid through Terzaghi's concept of effective stress[19,20,36] leads to the following relation (see Appendix C for derivation):

$$e_x = \frac{du}{dx} = \frac{1}{E_{oed}}p. \qquad (3)$$

where $u$ is the displacement in the $x$ direction and $E_{oed}$ is the Oedometric modulus[38,39]. Substituting equation (2) in (3) gives:

$$e_x = \frac{du}{dx} = \frac{1}{E_{oed}}\left(-\frac{p_{in}}{L}x + p_{in}\right). \qquad (4)$$

Under the condition that the inlet boundary is fixed ($u(x = 0) = 0$), we integrate (4) to derive a solution for the displacement at each location along the $x$ axis:

$$u(x) = -\frac{p_{in}}{2E_{oed}L}x^2 + \frac{p_{in}}{E_{oed}}x. \qquad (5)$$

applying equation (5) on the sample's edge ($x = L$) we can write:

$$p_{in} = 2E_{oed} \frac{u_{edge}}{L}. \qquad (6)$$

Equation (6) suggests that when the one-dimensional strain ($u_{edge}/L$) is caused by pore-pressure instead of external force, the sample's edge will displace as if the material is twice as stiffer than in a regular Oedometric test. This is due to the fact that in regular Oedometric test the stress is constant along the medium, while here the effective stress follows the pore pressure and hence decreases along the sample. Moreover, $2E_{oed}$ is exactly the slope of our pseudo stress-strain curve (figure 2a, red line). Hence, from the slope of the measured pseudo stress-strain curve we can directly calculate $E_{oed}$, without measuring $E$ and $\nu$ in a separate rheological test. Knowing the value of $E_{oed}$, we can use equation (5) to model the displacement in each point along our sample for all the inlet pressures and *without any fitting parameters* (figure 3, brown line). As can be seen in figure 3, the model agrees extremely well with the measured data. Knowing the value of $E_{oed}$, equation (3) can be used for calculating the strain along the sample (figure 3, purple line). Following the pore pressure, the strain decreases linearly along the sample. While the injection area is highly expanding, the outlet area hardly strains.

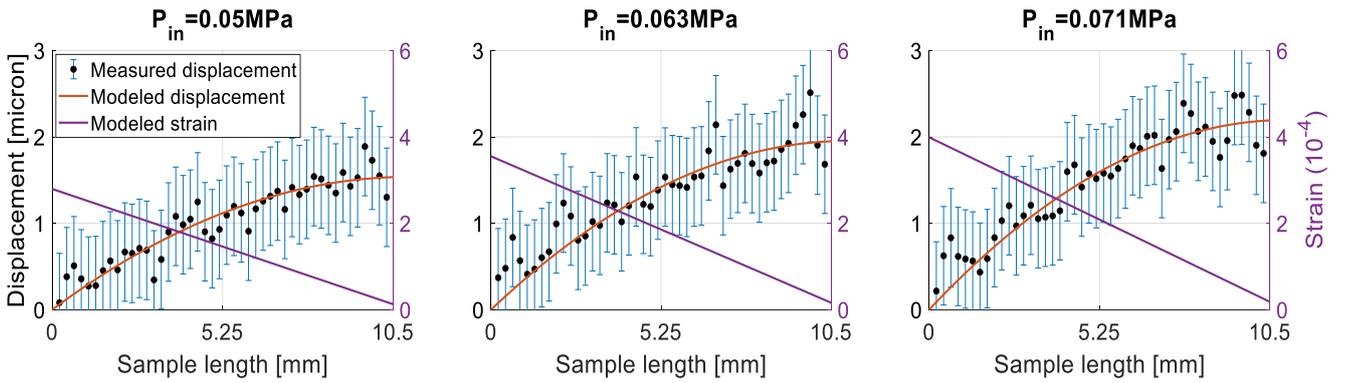

*Figure 3. Internal displacement along the sample. The model we derived based on the theory of poroelasticity (brown curve) predicts accurately the measured displacement (black dots), without any fitting parameters. Using the model we also calculated the internal strain (purple curve), which decreases linearly along the medium, following the pore pressure. While the injection area is strongly expanding the sample's far edge hardly strains.*

# Discussion

In this study we used a unique rock-like medium together with a novel experimental system to visualize the internal displacement field induced by pressurized flow through the porous medium. We showed that one dimensional fluid injection induces linear decreasing expansion, which lead to parabolic increase in the cumulative displacement along the medium, a phenomenon that we modeled accurately without any fitting parameters using the poroelastic theory. Hence, this study confirms and describes the applicability of the poroelastic theory in modeling injection induced elastic deformation of rocks. The important assumptions made in the classical work of Biot (1943)[18] : linearity and reversibility of strain-pressure relations and constant permeability due to small deformations, proved to be correct for the case of fluid injection described in our system. In addition, our results suggest that the medium is strained due to the absolute pore-pressure alone and not due to the drag force exerted on the medium by the flow. In future studies we intend to investigate under which conditions the drag force becomes a strain driving force.

The presented analytical model can be easily modified to adjust to the natural boundary conditions for modelling the actual surface uplift caused by poroelastic expansion due to fluid injection into the underground. This can be done by modifying the model dimensions into a 2D polar coordinate system to account for a circular uniform fluid injection into a confined aquifer. The Oedometric modulus we mentioned in our 1D model can be further used for the natural boundary conditions as the confined aquifer has one direction of free strain (i.e., the surface) and two other directions of limited strain (i.e., the underground), just like in Oedometric test. Although this work was focused on elastic deformation, our rock-like medium has the potential to explore the full range of elasto-plastic transition, which we will investigate in a subsequent study.

# Acknowledgments

We thank Alberto Guadagnini and Gabriele Dellavecchia for their insightful comments and discussions. We also thank Ludmila Abezgauz, Shaimaa Sulieman and Martin Stolar for their help in the experiments. A.B. acknowledge The Nancy and Stephen Grand Technion Energy Program (GTEP) for supporting this research. Y.E. and A.B. thank the support of ISF-NSFC (grant No. 3333/19)

# Supplemental material

Appendix A: Material properties

The grain size distribution of the porous medium used in this research was derived for the unconsolidated PMMA particles composing the medium (figure 4a) by a Dynamic light scattering analysis (DLS) (figure 4b). The mean grain size of the medium is 78 microns. The fluorescent beads (106-125 microns in diameter) are within the range of the initial distribution and fit within the tail of it.

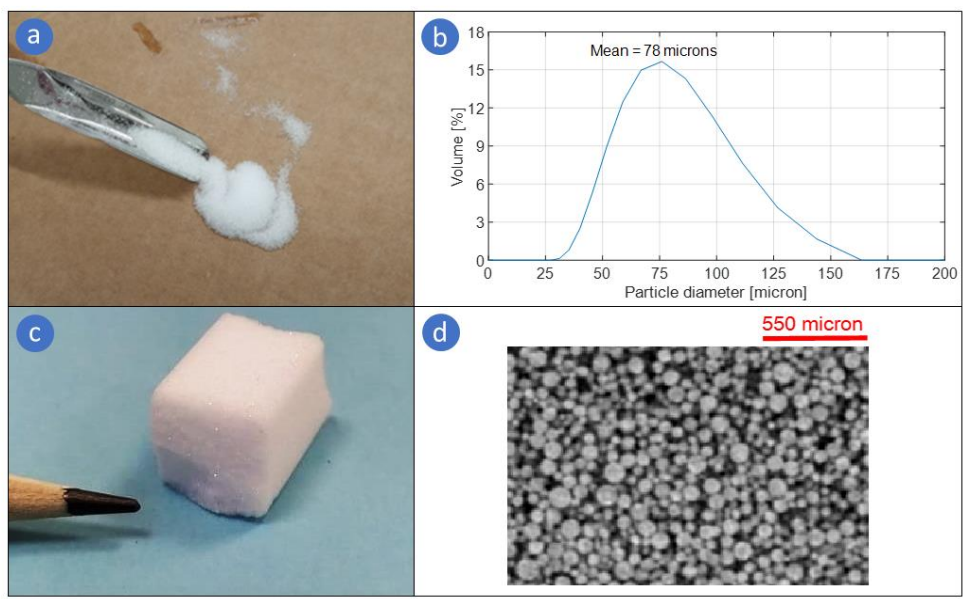

*Figure 4. Material description and properties. a: The PMMA beads composing the artificial rock, before sintering. b: DLS analysis of the non-sintered PMMA beads. c: The artificial rock, made by sintering the PMMA beads. d: Micro CT scan of the artificial rock (one plain).*

The sintered porous medium (figure 4c) has been scanned by micro-CT to study its internal structure (figure 4d, resolution: 1 pixel= 7.5 microns). For calculating the porosity, the whole scanned volume has been binarized through Matlab, using Otsu's method[40] and the porosity ($n = 0.43$) has been calculated as the ratio between the black pixels (voids) to the overall number of pixels in the volume. The porosity analysis has been verified in another test in which the consolidated porous medium is saturated with oil inside the flow cell. The weight of the oil reservoir is measured before the saturation and after it so the difference is exactly the weight of the oil inside the porous medium. Knowing the density of the oil, the pore fluid volume is calculated and divided by the measured volume of the medium to achieve the porosity. The porosity value that was calculated from this test was 0.42, almost the same as the porosity value from the CT test.

## Appendix B: PIV Analysis

The PIV calculates the mean displacement value for each 448X448 microns (32X32 pixels) along the sample, providing the whole internal displacement field for each inlet pressure applied (figure 5). This size of interrogation area was optimized by the software[33], with 224 microns (16 pixels) overlap between each area. We analyze the internal displacement along the sample (figure 3) by calculating the mean displacement of each column of interrogation units along the sample's $x$ axis. The mean displacement is correlated to the distance of the interrogation unit center from the sample's inlet, resulting in internal displacement values for each 224 microns along the sample.

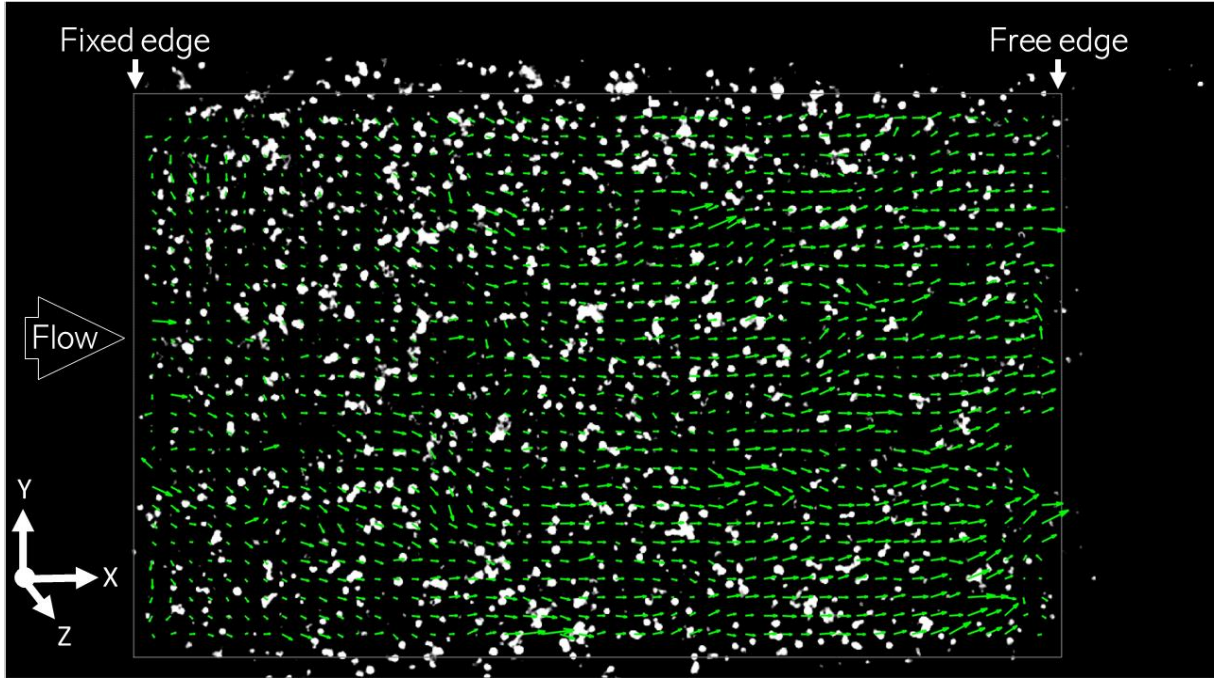

*Figure 5. Full PIVlab analysis[33]. The PIV analysis provides the whole internal displacement field of the sample for each pressure difference applied. The displacements are in the direction of the empty space at the sample's outlet. There is a clear increase in the magnitude of the displacement along the sample, indicating that the medium is expanding on account of the empty space and not moving as a rigid body.*

# Appendix C: Model derivation

The stress (*σ*) - strain (*e*) relations are governed by Hooke's law for an isotropic elastic body:

$$e_x = \frac{\sigma_x}{E} - \frac{\nu}{E}(\sigma_y + \sigma_z).$$

(7)

$$e_y = \frac{\sigma_y}{E} - \frac{\nu}{E}(\sigma_x + \sigma_z).$$

(8)

$$e_z = \frac{\sigma_z}{E} - \frac{\nu}{E}(\sigma_y + \sigma_x).$$

(9)

where $E$ is Young's modulus, and $\nu$ is Poisson's ratio.

To account for the presence of fluid in the medium, we will use Terzaghi's concept of effective stress[19,20,36], which coincides with Biot's theory of poroelasticity[18] for incompressible grains and fluid. Defining tension and expansion as positive:

$$\sigma' = \sigma_{ex} + p.$$

(10)

where $\sigma'$ is the effective stress, $\sigma_{ex}$ is external stress and $p$ is the pore-pressure.

For this set-up, there is no external stress in the $x$ direction, thus:

$$\sigma'_x = p.$$

(11)

Our boundary conditions dictate that $e_y = e_z = 0$, therefore, equation (8) in terms of effective stress can be written as:

$$0 = \frac{\sigma'_y}{E} - \frac{\nu}{E}(\sigma'_x + \sigma'_z).$$

(12)

assuming that $\sigma'_y = \sigma'_z$ we can write:

$$0 = \frac{\sigma'_y}{E} - \frac{\nu}{E}(\sigma'_x + \sigma'_y).$$

(13)

using equation (13), we can find $\sigma'_y$ (and $\sigma'_z$) as function of $\sigma'_x$:

$$\sigma'_y = \sigma'_z = \frac{\nu}{1-\nu}\sigma'_x.$$

(14)

substitution of equation (14) in (7) gives:

$$e_x = \frac{\sigma'_x}{E} - \frac{\nu}{E}\left(2\frac{\nu}{1-\nu}\sigma'_x\right).$$

(15)

equation (15) can be rewritten as:

$$e_x = \frac{1}{E}\left(1 - \frac{2v^2}{1-v}\right)\sigma'_x .$$

(16)

substituting equation (11) in (16) gives:

$$e_x = \frac{1}{E}\left(1 - \frac{2v^2}{1-v}\right)p .$$

(17)

the coefficient of the pore pressure is the inverse of the *Oedometric modulus*[38,39]. That is:

$$e_x = \frac{1}{E_{oed}}p .$$

(18)

where:

$$E_{eod} = \frac{E}{1 - \dfrac{2v^2}{1-v}} .$$

(19)